\documentclass[floats, prd, eqnum, showpacs, nofootinbib, 
twocolumn,
eqsecnum]{revtex4-1}

\usepackage{color,graphicx}
\usepackage{amsfonts}
\usepackage{amssymb}
\usepackage{comment}
\begin{document}

\title{Is there a trade-off relation between efficiency and power in a collisional Penrose process 
in an extreme Reissner-Nordstr\"{o}m spacetime?} 
\author{Naoki Tsukamoto${}^{1}$}\email{tsukamoto@rikkyo.ac.jp}

\affiliation{
${}^{1}$Department of General Science and Education, National Institute of Technology, Hachinohe College, Aomori 039-1192, Japan \\
}

\begin{abstract}
We investigate the power of a collisional Penrose process with an unbound energy extraction from an extreme Reissner-Nordstr\"{o}m black hole.
This process takes infinite time in a time coordinate at a constant radial coordinate outside of the black hole. 
For black holes as a power plant, the power of the process for an observer far away from the black hole can be useful.
We define the power as energy gain from the extreme Reissner-Nordstr\"{o}m black hole divided by the time interval of the process in a coordinate time;
we estimate the upper bound of the power in a near-horizon limit, while the efficiency of the process can be arbitrary large in that limit. 
Thus, we conclude that there is no trade-off relation between the efficiency and power in the collisional Penrose process in extreme Reissner-Nordstr\"{o}m spacetime.
\end{abstract}
\maketitle

\section{Introduction}
In 1969, Penrose suggested a process, called the Penrose process, to extract rotational energy from a Kerr black hole by 
dropping a particle with negative energy in the black hole~\cite{Penrose:1969pc}.
Denardo and Ruffini pointed out that the electromagnetic counterpart 
of the Penrose process occurs in Reissner-Nordstr\"{o}m spacetime~\cite{Denardo:1973pyo}, and 
Blandford and Znajek considered the electromagnetic extraction of rotational energy 
from the Kerr black hole in an astrophysical situation~\cite{Blandford:1977ds}.
It is considered that the electromagnetic extraction of energy from the black holes can explain high energy jets near them. 

Piran, Shaham, and Katz~\cite{Piran:1975} and T.~Piran and J.~Shaham~\cite{Piran:1977dm}
investigated a particle collision near the Kerr black hole. 
They pointed out that the center-of-mass energy of two particles can be arbitrarily high 
if the Kerr black hole has an extreme event horizon and one of two particles has a critical angular momentum.
Particle collision with infinite center-of-mass energy is called the Ba\~{n}ados-Silk-West process 
since it was rediscovered by Ba\~{n}ados, Silk, and West in 2009~\cite{Banados:2009pr}.

After rediscovering the BSW process, 
several authors critically investigated the process with arbitrarily high center-of-mass energy:
(i) The Kerr black hole cannot have an extreme rotation in an astrophysical situation~\cite{Berti:2009bk}, 
(ii) gravitational radiation and its backreaction constrain the center-of-mass energy for the particle collision~\cite{Berti:2009bk}, 
(iii) to obtain the unbounded center-of-mass energy, an infinite proper time of a falling particle to reach the extreme event horizon is required~\cite{Jacobson:2009zg}, and 
(iv) self-gravity of falling objects constrains the center-of-mass energy for the collision~\cite{Kimura:2010qy,Ogasawara:2018gni}.

Reference~\cite{Banados:2009pr} inspired several authors to investigate the details of the collisional Penrose process~\cite{Piran:1977dm}.
The upper limit of energy extraction by the collision Penrose process after the BSW collision near the 
extreme Kerr black hole is very modest~\cite{Bejger:2012yb,Harada:2012ap}. 
Schnittman found that a collisional Penrose process with energy gain can be more than $10$ times the energy of 
incident particles~\cite{Schnittman:2014zsa,Leiderschneider:2015kwa,Ogasawara:2015umo,Harada:2016eff,Zaslavskii:2016unn}.
A collisional Penrose process after the head-on collision of two particles near 
an extreme Kerr black hole~\cite{Berti:2014lva,Leiderschneider:2015ika,Zaslavskii:2015ema,Ogasawara:2016yfk},
collisional Penrose processes with spinning particles~\cite{Maeda:2018hfi,Okabayashi:2019wjs}, and 
collisional Penrose processes in wormhole spacetimes~\cite{Tsukamoto:2015hta,Zaslavskii:2018kix}
and in an overspinning Kerr spacetime~\cite{Patil:2015fua}
were investigated. 

Zaslavskii found the electromagnetic counterpart of BSW collision 
near an extreme charged Reissner-Nordstr\"{o}m black hole~\cite{Zaslavskii:2010aw}.
Zaslavskii pointed out that energy extraction from an extreme Reissner-Nordstr\"{o}m black hole 
in a collisional Penrose process after the BSW collision can be unbound~\cite{Zaslavskii:2012ax,Nemoto:2012cq} 
while the energy extraction from the extreme Kerr black hole is very modest~\footnote{Bhat, Dhurandhar, and Dadhich found 
no upper limit on the efficiency of a Penrose process with electromagnetic interaction in a Kerr-Newman black hole spacetime in Ref.~\cite{Dhurandhar_1985}.}.
The Reissner-Nordstr\"{o}m spacetime is more tractable than the Kerr black hole due to spherical symmetry of the spacetime.
A finite center-of-mass energy of BSW collisions of two shells including their self-gravity was shown in Ref.~\cite{Kimura:2010qy},
and an upper bound of energy extraction from the extreme Reissner-Nordstr\"{o}m black hole, 
by fully taking into account the self-gravity of the colliding shells, was obtained in Ref.~\cite{Nakao:2017xwe}~\footnote{Multiple BSW collisions 
and multiple Penrose processes in Reissner-Nordstr\"{o}m spacetime~\cite{Kokubu:2020jvd,Kokubu:2021cwj} and a BSW collision 
in higher-dimensional Reissner-Nordstr\"{o}m spacetime~\cite{Tsukamoto:2013dna} were also investigated.}.

In this paper, inspired by a trade-off relation between efficiency and power of a heat engine by Shiraishi, Saito, and Tasaki~\cite{Shiraishi},
we consider the power of the collisional Penrose process, which gives infinite efficiency in the extreme Reissner-Nordstr\"{o}m spacetime 
after the BSW collision. 
Is there a trade-off relation between efficiency and power of the collisional Penrose process? 
To answer this question, we define the power of the collisional Penrose process as 
the energy extraction divided by a coordinate time, and we discuss the maximum of the power in the process.

This paper is organized as follows. 
In Sec.~II, we review the motion of a charged particle in an extreme Reissner-Nordstr\"{o}m spacetime. 
In Sec.~III, we review the energy extraction from the extreme Reissner-Nordstr\"{o}m black hole in a collision Penrose process
and we investigate the power of the process.
In Sec.~IV, we conclude with our results.
In this paper, we use geometrical units in which the light speed and Newton's constant are unity.

\section{Motion of a charged particle in a Reissner-Nordstr\"{o}m spacetime}
A line element and a vector potential in a Reissner-Nordstr\"{o}m spacetime are expressed by
\begin{eqnarray}
&&ds^2=-f(r)dt^2+\frac{dr^2}{f(r)}+r^2(d\theta^2+\sin^2 \theta d \phi^2), \\
&&A_\mu dx^\mu = -\frac{Q}{r}dt, 
\end{eqnarray}
where $f(r)$ is given by 
\begin{eqnarray}
f(r) \equiv 1- \frac{2M}{r}+\frac{Q^2}{r^2},
\end{eqnarray}
and $Q$ and $M$ are electrical charge and mass, respectively.
We find a black hole spacetime with an event horizon at $r=r_{\mathrm{H}}\equiv M+\sqrt{M^2-Q^2}$ for $\left| Q \right| \leq M$ 
and a spacetime with naked singularity for $\left| Q \right| > M$.
We assume an extreme charge $Q=M>0$ since we are interested in a collisional Penrose process with an unbound energy extraction~\cite{Zaslavskii:2012ax,Nemoto:2012cq}.

From the Hamiltonian equation, the four-momentum $p^\mu$ of a particle with an electrical charge $q$ is expressed as 
\begin{eqnarray}\label{eq:Hamiltonian}
p^\mu
&=&\frac{\partial H}{\partial \pi_\mu} \nonumber\\
&=&\pi^\mu-qA^\mu,
\end{eqnarray}
where $H$ is the Hamiltonian of the charged particle given by
\begin{eqnarray}
H\equiv \frac{1}{2} g^{\mu \nu} (\pi_\mu-qA_\mu)(\pi_\nu-qA_\nu),
\end{eqnarray}
and $\pi^\mu$ is the canonical momentum of the charged particle conjugate to the coordinates $x^\mu$.
We assume that charged particles have vanishing angular momentum $L\equiv \pi_\phi=0$
and that they only move in a radial direction on an equatorial plane $\theta=\pi/2$.

From the $t$ component of the four-momentum $p^\mu = dx^\mu / d\lambda$, where $\lambda$ is an affine parameter, and from Eq.~(\ref{eq:Hamiltonian}), 
we get 
\begin{eqnarray}
\frac{dt}{d\lambda}= \frac{1}{f}\left( E-\frac{qM}{r} \right),
\end{eqnarray}
where $E\equiv -\pi_t$ is the conserved energy of the charged particle.
The particle should satisfy a forward-in-time condition $dt/d\lambda \geq 0$.
The condition is expressed by
\begin{eqnarray}
E-\frac{qM}{r}\geq 0
\end{eqnarray}
and it yields, for $r=r_{\mathrm{H}}=M$,
\begin{eqnarray}
E=q.
\end{eqnarray}
We call a charged particle with $E=q$ a critical particle.

From $g_{\mu\nu} p^\mu p^\nu=-m^2$, where $m$ is the mass of the charged particle, 
we obtain 
\begin{eqnarray}
\left( \frac{dr}{d\lambda} \right)^2+V(r)=0,
\end{eqnarray}
where $V(r)$ is the effective potential of the radial motion of the charged particle given by 
\begin{eqnarray}
V(r)\equiv -\left(E-\frac{qM}{r}\right)^2+m^2f.
\end{eqnarray}
The charged particle can exist only in a region where the effective potential $V(r)$ is nonpositive.
The radial component of the four-momentum of a particle can be written as $p^r=\sigma \sqrt{-V}$, where 
$\sigma=-1$ ($\sigma=1$) for an ingoing (outgoing) particle.

\section{Particle collision and energy extraction from a black hole}
In this section, we review energy extraction from the extreme Reissner-Nordstr\"{o}m black hole in a collision Penrose process~\cite{Zaslavskii:2012ax,Nemoto:2012cq}
as well as investigate power in this process.
We consider that particles $1$ and $2$ collide at $r=r_{\mathrm{c}}\equiv M(1+\epsilon)$, where $0<\epsilon \ll 1$, 
and particles $3$ and $4$ are produced after the collision.
We set $\sigma_1=\sigma_2=-1$.
Here and hereinafter, physical values with subscripts $1$, $2$, $3$, and $4$ denote physical values of particles $1$, $2$, $3$, and $4$, respectively.
The center-of-mass energy $E_\mathrm{CM}$ of particles $1$ and $2$ at the collision is given by
\begin{eqnarray}
E_\mathrm{CM}^2
&\equiv&-g_{\mu \nu}(p^\mu_1+p^\mu_2)(p^\nu_1+p^\nu_2) \nonumber\\
&=&m^2_1+m^2_2+\frac{2}{f(r_\mathrm{c})} \left[  \left( E_1-\frac{q_1M}{r_\mathrm{c}} \right)  \left( E_2-\frac{q_2M}{r_\mathrm{c}} \right) \right. \nonumber\\
&&\left. +\sqrt{-V_1(r_\mathrm{c})}\sqrt{-V_2(r_\mathrm{c})} \right].
\end{eqnarray}
If one of the particles is critical and the other is not critical, the center-of-mass energy diverges 
in a near-horizon limit $\epsilon \rightarrow 0$~\cite{Zaslavskii:2010aw}.
For simplicity, we assume that particle $1$ is critical, $E_1=q_1$, and particle~$2$ has no charge, $q_2=0$.
The center-of-mass energy is obtained as 
\begin{eqnarray}
E_\mathrm{CM}^2
&=&m^2_1+m^2_2+\frac{2(1+\epsilon)}{\epsilon} \left[  E_1 E_2 \right. \nonumber\\
&&\left. -\sqrt{E_1^2-m_1^2}\sqrt{E_2^2-m_2^2\left( \frac{\epsilon}{1+\epsilon} \right)^2} \right] \nonumber\\
&\sim&\frac{2A_1 E_2}{\epsilon},
\end{eqnarray}
where $A_1$ is defined by $A_1\equiv E_1-\sqrt{E_1^2-m_1^2}$.

The conservation law of the charges before and after the particle collision is expressed by 
\begin{eqnarray}\label{eq:charge}
q_1+q_2=q_3+q_4.
\end{eqnarray}
The conservation law of the four-momentum of the particles at the moment of the collision is expressed by
\begin{eqnarray}\label{eq:momentum}
p^\mu_1+p^\mu_2=p^\mu_3+p^\mu_4.
\end{eqnarray}
The $t$ component of the conservation laws of the four-momentum (\ref{eq:momentum}) and the charges (\ref{eq:charge}) gives
the conservation laws of conserved energy, 
\begin{eqnarray}\label{eq:energy}
E_1+E_2=E_3+E_4.
\end{eqnarray}

We assume that particle~$3$ is a near-critical particle with $q_3=E_3(1+\delta_3 \epsilon)$, where $0<\delta_3<1$,
and that particles~$3$ and $4$ are ingoing particles with $\sigma_3=\sigma_4=-1$ 
immediately after the creation of the particles~\cite{Zaslavskii:2012ax,Nemoto:2012cq}.
From the radial component of Eq.~(\ref{eq:momentum}), we obtain  
\begin{eqnarray}\label{eq:a}
A_1+E_3 (\delta_3-1)=-\sqrt{E_3^2(1-\delta_3)^2-m_3^2}.
\end{eqnarray}
From the square of Eq.~(\ref{eq:a}), we get
\begin{eqnarray}\label{eq:b}
\delta_3=1-\frac{m_3^2+A_1^2}{2A_1E_3}.
\end{eqnarray}

We assume that particle~$3$ is reflected at a turning point $r=r_-$, 
and we assume that $E_3 > m_3$ so that particle~$3$ goes to spatial infinity.
For simplicity, we also assume that particle~$4$ does not interact with particle~$3$ after their particle production and 
particle~$4$ falls into the event horizon of the black hole~\footnote{One may be concerned that particle~$4$ collides with particle~$3$. 
By taking nonzero conserved angular momenta of particles~$3$ and $4$ into account, the collision can be avoided.}.
From $V_3(r_-)=0$, we obtain the turning point $r=r_-$ as
\begin{eqnarray}
r_{-}
=M\left( 1 + \frac{E_3 \delta_3 \epsilon }{E_3 - m_3} \right).
\end{eqnarray}
From $r_{-} \leq r_{\mathrm{c}}$, the condition 
\begin{eqnarray}\label{eq:cond}
E_3 (1-\delta_3) \geq m_3
\end{eqnarray}
must be satisfied. 
If the inequality~(\ref{eq:cond}) holds,
the inside of the square root in (\ref{eq:a}) is positive, 
and the lower bound of the mass of particle $3$, 
\begin{eqnarray}
m_3 \geq A_1
\end{eqnarray}
must hold so that the left-hand side of Eq.~(\ref{eq:a}) is negative.
For simplicity, we assume $m_0\equiv m_1=m_2$ and $E_1=E_2\gtrsim m_0$.
As discussed in Ref.~\cite{Nemoto:2012cq}, 
the mass and conserved energy of particle $3$ can be  
$m_3 \sim E_\mathrm{CM} \sim m_0/\sqrt{\epsilon}$ and $E_3\sim m_0/\epsilon$, respectively.

The collisional Penrose process is expressed in Fig.~1: Particle~$1$ at an initial position $r=r_{\mathrm{i}}$ 
falls toward the extreme charged black hole and it collides with particle $2$ at $r=r_{\mathrm{c}}$ after the coordinate time of $\Delta t_1$. 
Particles $3$ and $4$ are produced from the collision. 
Particle $3$ is reflected at $r=r_-$, and it escapes to $r=r_{\mathrm{i}}$ after the coordinate time of $\Delta t_3$.
On the other hand, particle $4$ reaches the extreme event horizon at $r=r_{\mathrm{H}}$.
\begin{figure}[htbp]
\begin{center}
\includegraphics[width=85mm]{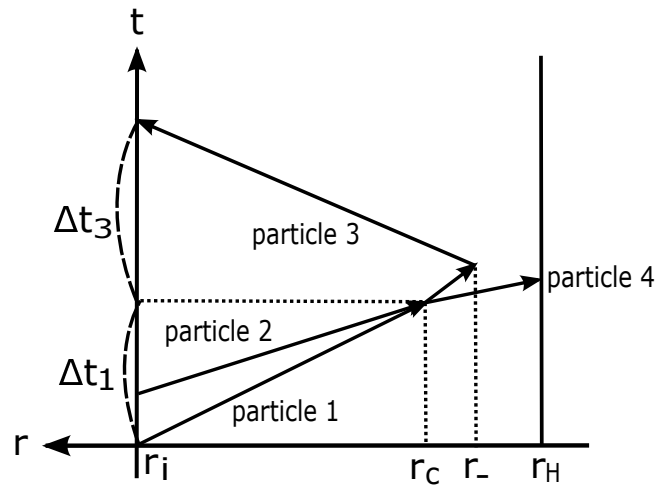}
\end{center}
\caption{Collisional Penrose process. Particle~$1$ with the critical condition $E_1=q_1$ at an initial position $r=r_{\mathrm{i}}$ 
moves toward the extreme Reissner-Nordstr\"{o}m black hole with $M=Q$ to collide with particle~$2$ with no charge, $q_2=0$, 
at $r=r_{\mathrm{c}}=M(1+\epsilon)$ after the coordinate time of $\Delta t_1$.
The collision between particles $1$ and $2$ produces particle $3$, with mass $m_3 \sim m_0/\sqrt{\epsilon}$ 
and conserved energy $E_3\sim m_0/\epsilon$, and particle $4$.
Particle $3$ is reflected at $r=r_{-}=M\left[ 1 + E_3 \delta_3 \epsilon/(E_3 - m_3)\right]$, 
and it reaches $r=r_{\mathrm{i}}$ after the coordinate time of $\Delta t_3$.
Particle $4$ falls into the extreme event horizon at $r=r_{\mathrm{H}}=M$.
}
\end{figure}
The time intervals $\Delta t_1$ and $\Delta t_3$ in the coordinate time are given by  
\begin{eqnarray}\label{eq:d2}
\Delta t_1
&=& -\frac{E_1}{\sqrt{E_1^2-m_0^2}}\int^{r_{\mathrm{c}}}_{r_{\mathrm{i}}} \frac{dr}{\left(1-\frac{M}{r}\right)^2} \nonumber\\
&\sim& \frac{E_1}{\sqrt{E_1^2-m_0^2}}\left( \frac{M}{\epsilon}+r_{\mathrm{i}} \right)
\end{eqnarray}
and
\begin{eqnarray}\label{eq:d3}
\Delta t_3 &=& -\int^{r_{-}}_{r_{\mathrm{c}}} \frac{E_3-\frac{q_3M}{r}}{\left(1-\frac{M}{r}\right)^2 \sqrt{-V_3}}dr \nonumber\\
&&+\int^{r_{\mathrm{i}}}_{r_{-}} \frac{E_3-\frac{q_3M}{r}}{\left(1-\frac{M}{r}\right)^2 \sqrt{-V_3}}dr \nonumber\\
&\sim& \left( \frac{2}{\delta_3}-2 \right) \frac{M}{\epsilon} +r_{\mathrm{i}},
\end{eqnarray}
respectively.

We define the power of the collisional Penrose process as 
$\Delta E/\Delta t$, where $\Delta E$ and $\Delta t$ are defined by $\Delta E \equiv E_3-E_1-E_2\sim m_0/\epsilon$ 
and $\Delta t \equiv \Delta t_1+\Delta t_3$, respectively.
From Eqs.~(\ref{eq:d2}) and (\ref{eq:d3}), the power of the collisional Penrose process is given by
\begin{eqnarray}\label{eq:power}
\frac{\Delta E}{\Delta t}\sim \frac{m_0}{\left( \frac{E_1}{\sqrt{E_1^2-m_0^2}} +\frac{2}{\delta_3}-2  \right)M+ \left( \frac{E_1}{\sqrt{E_1^2-m_0^2}} +1 \right)r_{\mathrm{i}}\epsilon} \nonumber\\
\end{eqnarray}
The power is plotted in Fig.~2.
\begin{figure}[htbp]
\begin{center}
\includegraphics[width=85mm]{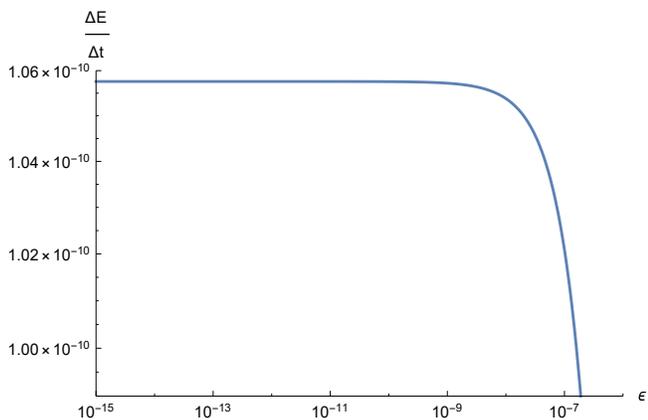}
\end{center}
\caption{Power of the collisional Penrose process given by Eq.~(\ref{eq:power}). We set $M=10$~km, $m_0=10^{-8}$~km, $E_1=1.1m_0$, and $r_{\mathrm{i}}=10^7$~km.}
\end{figure}
If $\epsilon \ll M/r_{\mathrm{i}}<1$ is satisfied, by using $E_3\sim m_0/\epsilon$, $m_3 \sim  m_0/\sqrt{\epsilon}$, and Eq.~(\ref{eq:b}), 
we get $\delta_3 \sim 1-m_0/(2A_1) \sim 1-(e_1+\sqrt{e_1^2-1})/2$, where $e_1$ is a specific conserved energy $e_1 \equiv E_1/m_0$ of particle~$1$, 
and the power is given by
\begin{eqnarray}\label{eq:max}
\frac{\Delta E}{\Delta t}
\sim \frac{1-e_1^2+(2-e_1)\sqrt{e_1^2-1}}{e_1^2+2e_1-2+e_1\sqrt{e_1^2-1}} \frac{m_0}{M}.
\end{eqnarray}
We notice that the collisional Penrose process has the upper bound of the power, 
the value of which is estimated to be as given in Eq.~(\ref{eq:max}) in the near-horizon limit $\epsilon \rightarrow 0$.

\section{Conclusion}
We have defined the power of the collisional Penrose process as 
the energy gain from the extreme Reissner-Nordstr\"{o}m black hole divided by the time interval of the process 
in a coordinate time.
The upper bound of the power can be estimated as given in Eq.~(\ref{eq:max}) in the near-horizon limit $\epsilon \rightarrow 0$. 
On the other hand, the efficiency of the collisional Penrose process defined by $\eta\equiv E_3/(E_1+E_2)$ 
has an arbitrarily high value $\eta \sim 1/(2\epsilon)$ in the near-horizon limit $\epsilon \rightarrow 0$ under our assumptions as discussed in Ref.~\cite{Nemoto:2012cq}.
Therefore, we conclude that there is no trade-off relation between efficiency and power in the collisional Penrose process in the extreme Reissner-Nordstr\"{o}m spacetime
under our treatment.
One may, however, find a trade-off relation between efficiency and power if the effect of the self-gravity of falling particles is taken into account.
The power of the collisional Penrose process including self-gravity of the particles is left for future work.

We now comment on the upper bound of the power in the collisional Penrose process. 
Zaslavskii categorized the scenarios of the collisional Penrose process with a single collision in the vicinity of the extreme event horizon 
to find unbounded energy extraction~\cite{Zaslavskii:2012ax}. 
All the scenarios, except for the one shown on Fig.~1, give vanishing power of the process in the near-horizon limit. 
However, one may find greater power than Eq.~(\ref{eq:power}) if 
some of the conditions on the extreme charge of the black hole, the critical charge of particles, and the near-horizon collision are violated.
Notice also that the upper bound of the power~(\ref{eq:power}) is not exact due to approximations and simplifications. 
In addition, the power of the collisional Penrose processes can be enhanced 
if we consider the collision of spinning particles~\cite{Armaza:2015eha,Maeda:2018hfi,Okabayashi:2019wjs}.
On the other hand, if the self-gravity of falling objects is taken into account, the power would be suppressed~\cite{Nakao:2017xwe}.  

The secondary collisional Penrose process of a head-on collision~\cite{Piran:1977dm,Zaslavskii:2013nra,Tsukamoto:2019ihj} 
between particle~$3$, which is reflected near the black hole, and an additional falling particle may increase the power of the total process
even though the self-gravity of particles~$3$ and $4$ would decrease the power in the secondary collision.
Investigating the details of multiple collisional processes~\cite{Kokubu:2020jvd,Kokubu:2021cwj} 
can also be an interesting future work to find a larger power than Eq.~(\ref{eq:power}).
  
One may consider that the definition of the power of the process using the proper time of the particles 
is more desirable than one using the coordinate time. 
The effect of the difference of the definitions on the upper limit of the power is also left to future work.

\section*{Acknowledgements}
The author thanks an anonymous referee for useful comments.

\end{document}